\documentstyle[twoside,fleqn,espcrc2,epsf]{article}

\newcommand{\pspiccie}[1]{
\centerline{\setlength\epsfxsize{55mm}\epsfbox{#1}}}
\newcommand{\oneszero}{\mbox{}^1S_0}
\newcommand{\onepone}{\mbox{}^1P_1}
\newcommand{\threesone}{\mbox{}^3S_1}
\newcommand{\threep}[1]{\mbox{}^3P_{#1}}
\newcommand{\GeV}{\mbox{GeV}}

\newcommand{\tabl}[1]{table \ref{tab:#1}}
\newcommand{\fig}[1]{figure \ref{fig:#1}}

\title{ 
\vskip-2truecm
{\normalsize DAMTP--97-91}
\vskip1truecm
A study of the quenched $\overline{b}c$ mass spectrum at 
$\beta$=6.2\thanks{Poster presented by Hugh Shanahan}}

\author{C.T.H.~Davies\address{
Institute for Theoretical Physics,
University of California at Santa Barbara,\\
Santa Barbara, CA 93106-4030,
USA}, A.J.~Lidsey\address{
Dept. of Electronic and Electrical Engineering,
University of Leeds,\\
Leeds,  LS2 9JT, England, U.K.},
H.P.~Shanahan\address{ DAMTP, 21 Silver St., University of Cambridge,\\
Cambridge, CB3 9EW, England, U.K.}
(UKQCD Collaboration) }

\begin{document}

\begin{abstract}
We present an analysis of the mass spectrum of heavy quarkonia with non--degenerate quark 
masses. The heavier (bottom)
valence quark  is treated in a non--relativistic fashion  and the other (charm)  is a 
relativistic Wilson--like
quark using the improved SW action. Such states provide and 
interesting probe between the
relativistic B meson states and the non-relativistic bottomonium states.
\end{abstract}

\maketitle

\section{Introduction}

The mass spectrum of quarkonia with valence bottom and charm quarks are
interesting in a number of aspects. Though only a handful of 
events have been detected, upcoming B-factories should produce
copious amounts of them. Hence, we are put in the interesting 
circumstance of making a {\it pre}-diction rather than a {\it post}-diction
for non--exotic hadron masses. As they  interpolate the 
non-relativistic bottomonium states and fully relativistic 
heavy--light states such as the $B$ and $B_s$ they provide an
interesting probe of QCD dynamics. 
Of course, the calculation of these masses are also 
a first step in the calculation of decay constants and semi-leptonic widths,
and these are of interest phenomenologically as  
an accurate calculation of $f_{Bc}$ is crucial to obtaining as accurate
as possible a result for $f_B$ from LEP data \cite{fbc}
and the semi--leptonic decay $B_c \rightarrow J/\psi l \overline{\nu}$
will give another estimate of the matrix element $|V_{\mathrm{cb}}|$,
which will be seen at the LHC-B \cite{vcb}. 

\section{Lattice method }
Previous quenched lattice calculations of the 
$\overline{b}c$ spectrum  \cite{ajl,seyong-kim} has 
been carried out at $\beta=5.7$, where both valence quarks are  simulated
using the non-relativistic formalism \cite{NRQCD}.
On such coarse lattices, $am_b$ and $am_c >1$ and the effective Green's functions
for both quark masses 
can be calculated without introducing large radiative corrections to the
hamiltonian. 
However, it has been demonstrated  that for such lattice spacings  the charmonium spectrum
is not reproduced very accurately.
For quenched gluon fields at  $\beta=6.2$, quenched one can still
use the non-relativistic approach for the $b$ quarks
and the relativistic formalism can now be used for treating the $c$ quark.
As one expects the dominant errors to be a function of $am_c$, which still
approaches  1, we use the
Sheikholeslami--Wohlert action \cite{SW}, using 
a mean-field estimate of the clover coefficient
$c_{\mathrm{sw}}$. While not eliminating all errors up to 
${\cal O}(a^2)$, this mean field estimate approach does reduce
errors significantly \cite{beta_5_7}.

\section{Computational Details}
On a $24^3 \times 48$ lattice at $\beta=6.2$, corresponding to a lattice spacing
of approximately $2.7 \; \GeV$, two point functions were constructed
for the following states; $(1,2)\oneszero$, $(1,2)\threesone$, $\onepone$ and 
$\threep{0,1,2}$ (using the notation $(R) \mbox{}^{2s + 1}L_J $, where $R$ is the radial
excitation). The operators were calculated using Coulomb gauge fixed smeared 
sources\cite{ajl}. 
In order to maximise the statistics, the non-relativistic Green's functions
were calculated forwards and backwards in time, so that as many of the componants
of the relativistic propagator could be used. 
\begin{table*}[hbt]
\setlength{\tabcolsep}{1.5pc}
\caption{Mass parameters used for states}
\label{tab:masses}
\begin{tabular}{c|c|c}
\hline
\hline
$\kappa$ & $am_b$ & \# Configs. \\
\hline
$0.132$ & $1.22$ & 100 \\
$0.126$ & $1.22$ & 100 \\
$0.126$ & $2.44$ & 100 \\
\hline
\hline
\end{tabular}
\end{table*}
In \tabl{masses} the quark mass parameters and the number of 
configurations are listed. We note that $\kappa=0.126$ and 
$am_b=1.22$ correspond roughly
to charm and bottom quark masses 
from studies of charmonium and bottomonium systems respectively.
The NRQCD hamiltonian was corrected to ${\cal O}(mv^4)$,
using tadpole improved estimates of the relevant coefficients.    
As the configurations are quenched, the lattice spacing should be determined from 
one of the $\overline{b}c$ observables. 
As the  splitting $\delta m_{P,S} = \onepone - \mbox{}^1\overline{S}$ 
for charmonium and bottomonium only changes 
by a few MeV, we took  an average of these splittings and
hence assumed that 
$\delta m_{P,S}(\overline{b}c) = 0.4545 \;\GeV$.
\subsection{Mixing}
One expects that the $\onepone$ and $\threep{1}$ states 
to mix (this has been observed at $\beta=5.7$  \cite{ajl}
and with two relativistic propagators at higher $\beta$'s 
with different
masses around that of charm). 
Therefore we calculated the following matrix of 
correlators 
\begin{eqnarray*}
\left( \begin{array}{c c}
< \onepone \onepone^\dagger > &  < \onepone \threep{1}^\dagger > \\
< \threep{1} \onepone^\dagger > &  < \threep{1} \threep{1}^\dagger > 
\end{array}\right) \; \; .
\end{eqnarray*}
However, for all three combinations of quark masses,
the off--diagonal elements of this matrix were very noisy 
and were consistent with zero.

\section{Results}
\begin{figure}[htb]
\pspiccie{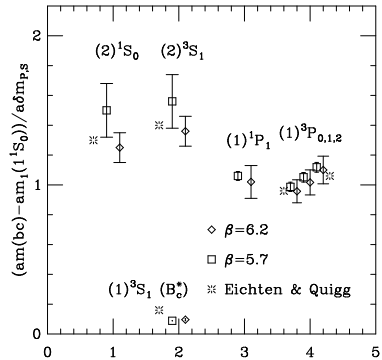}
\caption{  A comparison of the results at $\beta=5.7$\cite{ajl}, $\beta=6.2$ 
($\kappa=0.126$, $am_b=1.22$) and a potential model calculation. 
The labels on the horizontal axis are arbitrary.  }
\label{fig:com_vs_beta}
\end{figure}

In  \fig{com_vs_beta} the results for the $\overline{b}c$ spectrum
using the the best estimates for the bottom and charm quark masses
are compared with the previous lattice calculation at $\beta=5.7$ \cite{ajl}
and a potential model calculation  \cite{eichten}.
The $R=1$ states at both $\beta$'s are consistent with each other. 
For the $S$-wave states, where we have a measurement of the $R=2$
states, the agreement is to within two standard deviations, albeit with 
large errors. The potential model calculation   agrees with 
the lattice results except for the hyperfine 
splitting.  
It is not clear whether this difference is due any systematic
error in the potential model or by a choice of the scale
using $\delta m_{P,S}$. 

\begin{figure}[htb]
\pspiccie{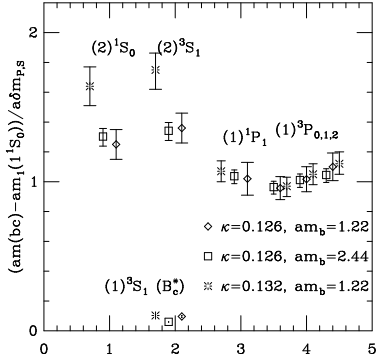}
\caption{ A comparison of the results at $\beta=6.2$ for 
the different combinations of quark mass. 
The labels on the horizontal axis are arbitrary. }
\label{fig:comp_vs_mass}
\end{figure}

\begin{figure}[htb]
\pspiccie{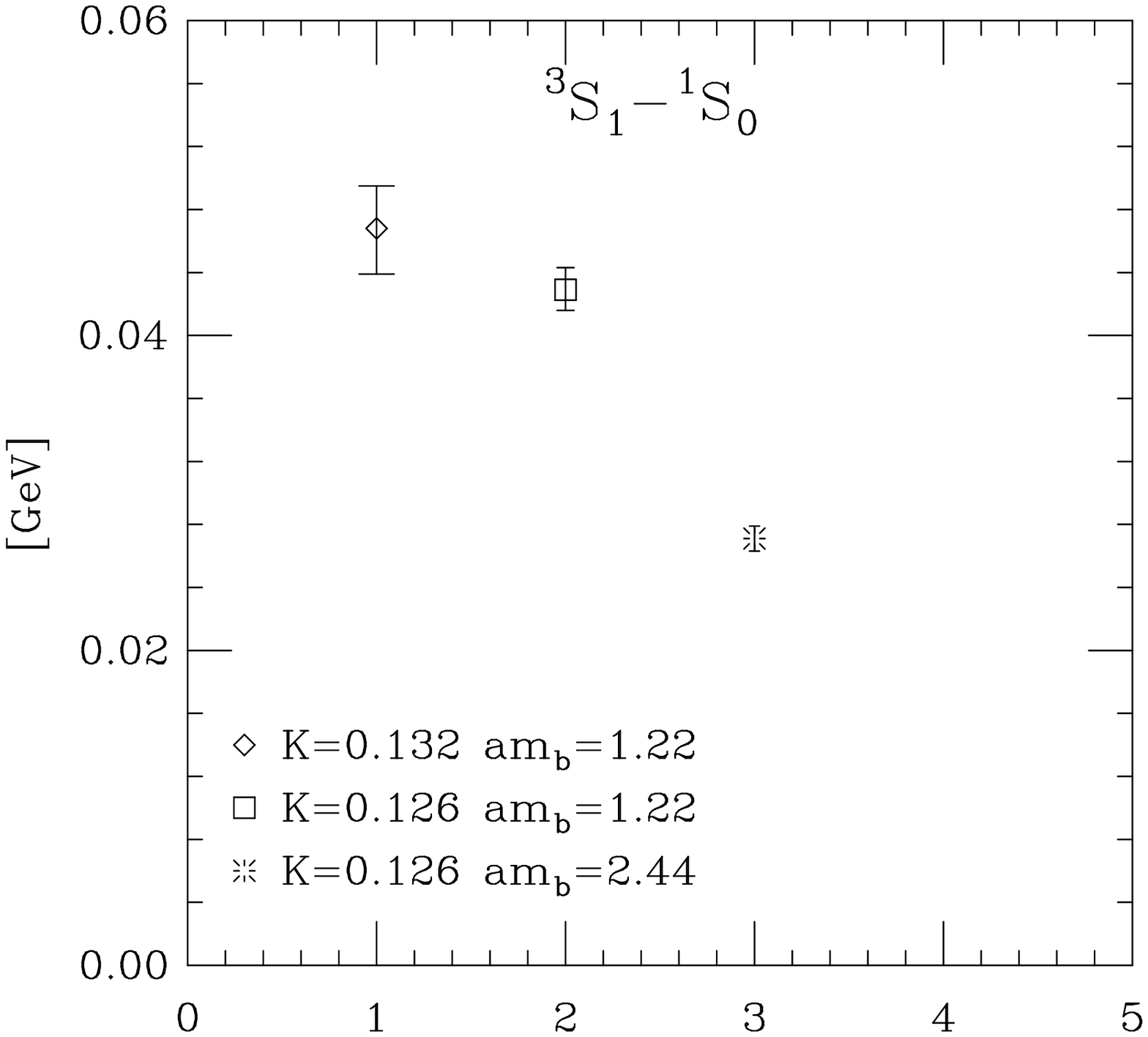}
\caption{ A comparison of the 
$\threesone - \oneszero$ splitting  at $\beta=6.2$ for 
the different combinations of quark mass. 
The labels on the horizontal axis are arbitrary. }
\label{fig:threesone_mass}
\end{figure}

\begin{figure}[htb]
\pspiccie{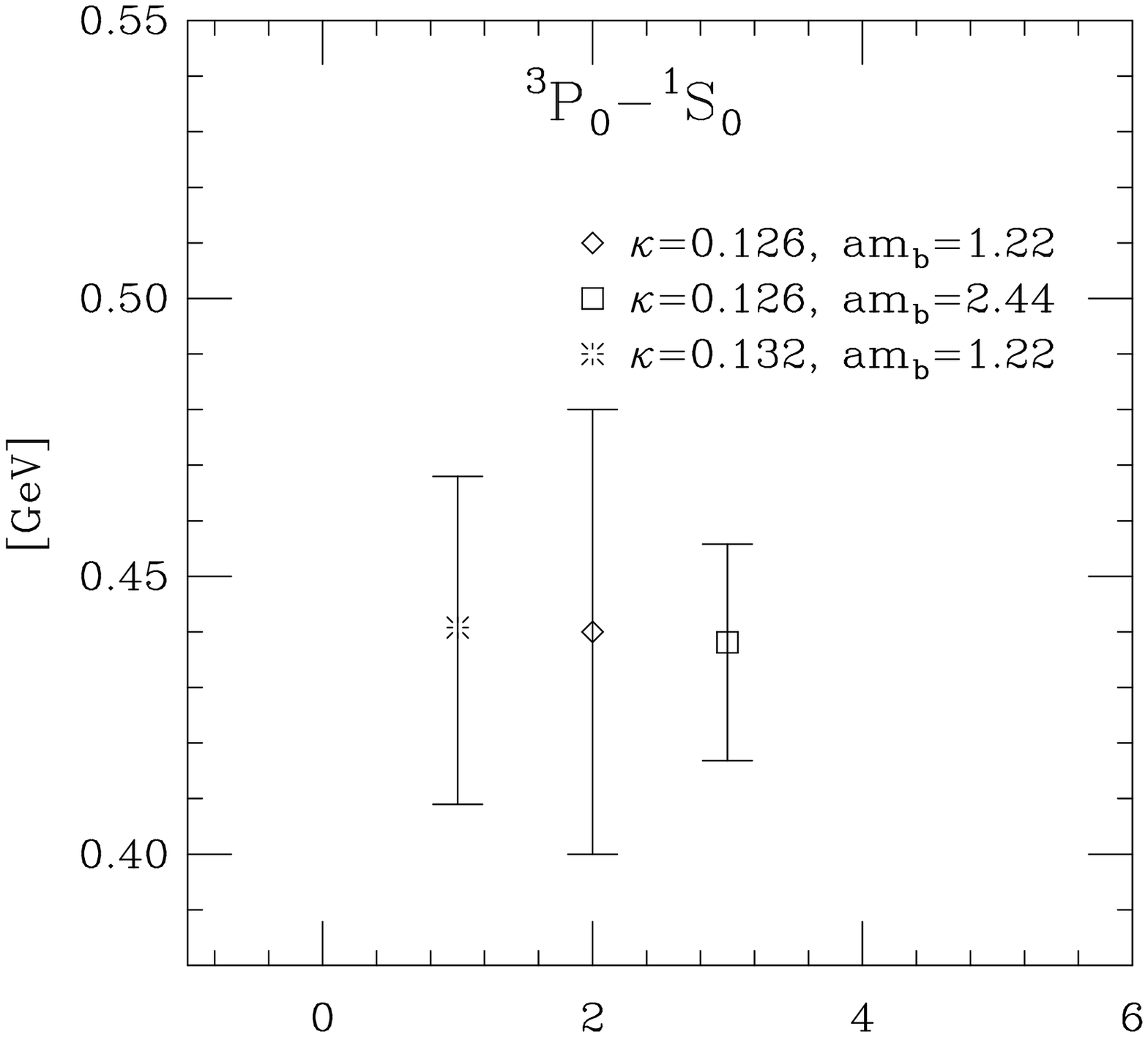}
\caption{ A comparison of the 
$\threep{0} - \oneszero$ splitting  at $\beta=6.2$ for 
the different combinations of quark mass. 
The labels on the horizontal axis are arbitrary. }
\label{fig:threepone_oneszero_mass}
\end{figure}

In \fig{comp_vs_mass}  the results for the $\overline{b}c$ spectrum
for different combinations of quark masses at $\beta=6.2$ are shown.
Once again. there is a general broad agreement. For the splittings,
we find that the splittings behave as expected. 
For example, in \fig{threesone_mass} we see that 
as the $b$ quark mass is increased, and the 
system behaves more like a heavy-light $bd$ state
the splitting tends to zero. Likewise, the fine
splitting (in \fig{threepone_oneszero_mass}) is 
constant, indicating that $\delta m_{P,S}$ should be
insensitive to variations in the quark masses.

\section{Conclusions}
It is very promising that the quenched lattice results at two different
lattice spacings are in good agreement with each  other, despite 
potential concerns over the size of the discretisation error
at $\beta=5.7$. The broad agreement with potential models is also
quite interesting as it is not clear that a semi-relativistic system
such as this could be modelled with a simple potential.
Given the amount of  research that has been done on 
the $\overline{b}b$ and $\overline{b}(u,d)$ spectrum,
it would be  interesting to see how the splittings of these
states vary as the masses of the quarks are changed.

The absence of mixing between the $\threep{1}$ and $\onepone$
states at $\beta=6.2$ is quite strange, given its presence at
$\beta=5.7$. It is of course possible that the mixing is a strong function
of the lattice spacing and that the mixing is very small in the continuum.
This would give a qualitatively different result from potential models.
In order to verify if this is true,  the calculation should be 
repeated using some other set of operators, defined in a different gauge
or a gauge invariant fashion.

\section{Acknowledgements}
The numerical work for this presentation was carried out on a Cray J90
at the EPCC in Edinburgh. HPS is supported by a grant from
the Leverhulme foundation.

\end{document}